\documentclass[aps,prl,nofootinbib,showpacs,preprint,floatfix]{revtex4}
\usepackage{latexsym,epsfig,graphicx}

\newcommand{\vs}[1]{\rule[- #1 mm]{0mm}{#1 mm}}
\newskip\humongous \humongous=0pt plus 1000pt minus 1000pt

\newif\ifdtup

\newcommand{\eq}{\vs{2}\begin{equation}}
\newcommand{\be}{\vs{2}\begin{equation}}
\newcommand{\en}{\\[2mm]\end{equation}}
\newcommand{\bea}{\begin{eqnarray}}
\newcommand{\ena}{\end{eqnarray}}
     
\newcommand{\lapprox}{%
\mathrel{%
\setbox0=\hbox{$<$}
\raise0.6ex\copy0\kern-\wd0
\lower0.65ex\hbox{$\sim$}
}}
\newcommand{\gapprox}{%
\mathrel{%
\setbox0=\hbox{$>$}
\raise0.6ex\copy0\kern-\wd0
\lower0.65ex\hbox{$\sim$}
}}

\begin{document}
\bibliographystyle{plain}
\begin{titlepage}
\begin{flushright}
{
UG-FT-134/02 \\
CAFPE-4/02 \\
June 2002\\}
\end{flushright}

\begin{center}
{\bf{\Large{\bf{Forward-backward asymmetries of lepton pairs 
in events with a large transverse momentum jet at hadron colliders}}}}
\\[0.5cm]
\indent
{\bf F. del Aguila$^a$, Ll. Ametller$^b$ and P. Talavera$^c$}
\\[0.5cm]
{$^a$ {\it Departamento de F{\'\i}sica Te\'orica y del Cosmos 
and Centro Andaluz de F{\'\i}sica de Part{\'\i}culas Elementales (CAFPE),
Universidad de Granada, 
E-18071 Granada, Spain.}\\[.1cm]
$^b$ {\it Departament de F{\'\i}sica i Enginyeria Nuclear,
Universitat Polit\`ecnica de Catalunya,\\
E-08034 Barcelona, Spain.}\\[.1cm]
$^c$ {\it Departament d'Estructura i Constituents de la Mat\`eria,
Universitat de Barcelona,\\
E-08028 Barcelona, Spain.}
}
\\[0.5cm]
\vfill
{\bf Abstract} \\
\end{center}
{We discuss forward-backward charge asymmetries for lepton pair 
production in association with a large transverse momentum jet 
at hadron colliders. 
The measurement of the lepton charge asymmetry relative to the 
jet direction ${\rm A}_{\rm FB}^j$ gives a new determination of the 
effective weak mixing angle 
$\sin ^2 \theta _{\rm eff}^{\rm lept}(M_Z^2)$ 
with in principle a statistical precision after cuts of 
$ 10^{-3}\ (8\times 10^{-3})$ at LHC (Tevatron),
due to the large cross section of the process with initial gluons. 
The identification of $b$ jets also allows for the 
measurement of the bottom quark $Z$ asymmetry ${\rm A}_{\rm FB}^b$ 
at hadron colliders although 
with a lower precision than at LEP, 
the resulting statistical precision for 
$\sin ^2 \theta _{\rm eff}^{\rm lept}(M_Z^2)$ 
being $\sim 9\times 10^{-4}$ ($2\times 10^{-2}$ at Tevatron).   
}
\noindent
\vfill
\begin{center}
 \\[0.2mm]
{\bf Keywords:} 
\begin{minipage}[t]{12cm}
Hadron-induced high- and super-high-energy interactions,
Gauge bosons. 
\end{minipage}
\end{center}
\end{titlepage}

\renewcommand{\thefigure}{\arabic{figure}}


The possibility of using hadron colliders to perform precision 
tests of the electroweak Standard Model (SM) is a challenge for 
the Fermilab Tevatron and the CERN Large Hadron Collider (LHC) 
experiments \cite{Tevatron,Haywood00}. 
Indeed, the large neutral gauge boson production cross section 
can allow for a precise determination of the effective 
weak mixing angle 
$\sin ^2\theta _{\rm eff}^{\rm lept}$,
the optimum observable being the forward-backward charge asymmetry 
of lepton pairs ${\rm A}_{\rm FB}$ in the Drell-Yan process 
$q\bar q \rightarrow \gamma, Z \rightarrow l^-l^+$, with 
$l=e$ or $\mu$ \cite{Rosner89,Fischer95}.
The Collider Detector Facility (CDF) Collaboration has reported 
a measurement of 
${\rm A}_{\rm FB} = 0.070 \pm 0.016$ for $e^-e^+$ pair 
invariant masses between 75 and 105 GeV at the Fermilab 
Tevatron Run I \cite{Abe96}. 
The expected precision to be reached at Run II with an integrated 
luminosity of 10 fb$^{-1}$ has been estimated to be 
$\sim 0.1 \%$, corresponding to a precision for 
$\sin ^2\theta _{\rm eff}^{\rm lept}$ of $\sim 0.05 \%$ 
\cite{Baur98}. At LHC with an integrated luminosity of 
100 fb$^{-1}$ the asymmetry precision will be further improved 
by a factor 
$\sim 6$ and the weak mixing angle one by a factor 
$\sim 3$. This is comparable 
to the global fit precision at LEP, but for instance a factor 
$\sim 2$ better than the weak mixing angle precision obtained 
from the bottom forward-backward asymmetry at the Z pole alone 
\cite{LEP02}.

The associated production 
of a neutral gauge boson $V=\gamma ,Z$ (with 
$V\rightarrow l^-l^+$) and a jet has also a large cross 
section, especially at LHC, thus can also allow 
for a precise determination of the effective weak mixing angle. 
This Next to Leading Order (NLO) correction to $V$ production 
is a genuine new process when the detection of the extra jet 
is required. In particular, gluons can be also initial states, 
and the large gluon content of the proton at high energy 
tends to make the $V$ and $Vj$ production cross sections of 
similar size. A neutral gauge boson with an accompanying jet 
is produced at tree level by 
$q\bar q$ and $g\stackrel {(-)}{q}$ collisions, amounting the 
latter to $\sim 83\ \%$ ($\sim 48\ \%$) of the total $Vj$ cross 
section at LHC, $\sqrt {s} = 14\, {\rm TeV}$ 
(Tevatron, $\sqrt {s} = 2\, {\rm TeV}$), for
the cuts below. 
In this Letter we point out that the forward-backward 
charge asymmetry of the lepton pairs can be measured 
in this process either relative 
to a direction fixed by the initial state 
${\rm A}_{\rm FB}$ as in the Drell-Yan 
case, or relative to the final jet direction 
${\rm A}_{\rm FB}^j$. 
The former is adapted to obtain the asymmetry 
from the events 
$q \bar q \rightarrow Vg \rightarrow l^-l^+g$, and the latter 
from $g \stackrel {(-)}{q} \rightarrow V \stackrel {(-)}{q} 
\rightarrow l^-l^+ \stackrel {(-)}{q}$.
Both asymmetries give similar precision for 
$\sin ^2\theta _{\rm eff}^{\rm lept}$ at LHC but not at 
Tevatron, where ${\rm A}_{\rm FB}$ gives a precision almost 
one order of magnitude higher. However, 
${\rm A}_{\rm FB}^j$ also 
allows for the measurement of flavour asymmetries. Thus, if 
we require the final jet to be a $b$ quark, we can make a 
new measurement of ${\rm A}_{\rm FB}^b$. 
This is especially interesting given 
its observed deviation at the $Z$ pole from the SM prediction, 
$2.9\ \sigma$ \cite{LEP02}. 
Although the precision in principle expected at LHC, 
$8.9\times 10^{-4}$, is lower than the one reported by LEP, 
$3.1\times 10^{-4}$, 
it is similar to the difference between the central 
values of $\sin ^2\theta _{\rm eff}^{\rm lept}$ 
resulting from ${\rm A}_{\rm FB}^b$ at the $Z$ pole
and the global fit to all data.

In the following we discuss these asymmetries and estimate 
the expected statistical precision at LHC and Tevatron. 
Here we present tree level results with no detailed detector 
simulation, although any attempt to fit real data demands 
including electromagnetic and strong as well as electroweak 
radiative corrections \cite{Baur98}. Although large, these 
corrections, which are expected to modify the predicted 
asymmetries appreciably, are NLO. 
Particularly worrisome is a priori the production of a neutral 
gauge boson with two jets.  
The phase space region where one of them is too soft or 
collinear dominates the total cross section. The corresponding
logarithmic behaviour then compensates for the corresponding
extra $\alpha _s$ suppression factor.
However, 
this leading contribution must be included in the 
Parton Distribution Functions (PDF) and then subtracted 
from the $Vjj$ cross section to avoid double counting, 
the resulting correction being actually NLO 
\cite{Catani96}. This process is also further enhanced 
at high energy for it has gluon fusion contributions. 
However, at LHC energies these are still smaller than the $Vj$ 
cross section. A detailed calculation of the NLO corrections 
is in progress.
The simulation of the 
experimental set up is also an essential ingredient to 
describe the observed asymmetries. We will try to mimic the 
experimental conditions in our parton calculation, 
but a real simulation is eventually needed.  
   
In Drell-Yan production of lepton pairs the forward-backward 
charge asymmetry has to be measured relative to the 
initial quark direction. In $p\bar p$ collisions this is 
identified with the direction of the proton 
because it has more quarks than antiquarks.       
\be
{\rm A}_{\rm FB} = \frac {F-B}{F+B}
\en 
with
\be
F = \int _0^1 \frac{{\rm d}\sigma}
{{\rm d}\cos \theta _{\rm CS}}
{\rm d}\cos \theta _{\rm CS}, \,\,\,  
B = \int _{-1}^0 \frac{{\rm d}\sigma}
{{\rm d}\cos \theta _{\rm CS}}
{\rm d}\cos \theta _{\rm CS} 
\label{asymmetry}
\en 
and
\be
\cos \theta _{\rm CS}=
\frac{2 (p_z^{l^-}E^{l^+}-p_z^{l^+}E^{l^-})}
{\sqrt {(p^{l^-} + p^{l^+})^2}
\sqrt {(p^{l^-} + p^{l^+})^2 + (p^{l^-}_T+p^{l^+}_T)^2 }},
\en 
where $\theta _{\rm CS}$ is the Collins-Soper angle 
\cite{Collins77}. The four-momenta are measured in the 
laboratory frame and
$p^{\mu}_T \equiv (0,p_x,p_y,0)$.
In $pp$ colliders the quark direction is fixed by the 
rapidity of the lepton pair. This implies defining 
$\cos \theta _{CS}$ with an extra sign factor 
$\frac{|p_z^{l^-}+p_z^{l^+}|}{p_z^{l^-}+p_z^{l^+}}$.
In $Vj$ production 
one can use 
the same asymmetry ${\rm A}_{\rm FB}$ 
or define a new one relative to the 
final jet, ${\rm A}_{\rm FB}^j$. In this last case the 
corresponding angle $\theta $ in Eq. (\ref{asymmetry}) 
is defined for $pp$ collisions as the angle between 
$l^-$ and the direction opposite to the jet in the 
$l^-l^+$ rest frame,
\be 
\cos \theta  = 
\frac{(p^{l^-}-p^{l^+})\cdot p^j}
{(p^{l^-}+p^{l^+})\cdot p^j}. 
\label{costheta} 
\en
The corresponding asymmetry which is suited to $g\stackrel {(-)}{q}$ 
collisions does not vanish 
because the proton contains many 
more quarks than antiquarks.
However, in $p\bar p$ colliders there are produced as 
many quarks as antiquarks and this asymmetry vanishes unless 
some difference is made between them.  
Hence, $\cos \theta $ is defined with an extra sign factor 
$\frac{|p_z|}{p_z}$, $p = p^{l^-} + p^{l^+} + p^j$,
which corresponds to 
assume that the largest rapidity parton is a (anti)quark 
if it is along the (anti)proton direction.
Besides, $Vj$ events also allow for measuring a 
flavour asymmetry if the final jet is identified and its charge 
determined, as in the case of $Vb$ production and 
${\rm A}_{\rm FB}^b$. 
For these events ${\rm A}_{\rm FB}$ is less significant. 
In order to obtain ${\rm A}_{\rm FB}^b$ in $pp$ 
or $p\bar p$ colliders one must use $\cos \theta$ in 
Eq. (\ref{costheta}) but multiplied by 
a $+(-)$ sign for b (anti)quarks, $-{\rm sign}(Q_b)$ 
with $Q_b$ the $b$ charge. 

Let us present our numerical results for 
$l^-l^+j$ and $l^-l^+b$ at LHC and Tevatron in turn.
We work in the effective Born 
approximation \cite{Haywood00} and use the MRST parton 
distribution functions \cite{Martin99}. 
The K factors for LHC and Tevatron,  
1.1 and 1.2, respectively \cite{Hamberg91}, are not included. 
Otherwise they would slightly improve our statistical 
precision estimates. 
Besides, we only count electron pairs. For muons the 
main differences would be the pseudorapidity 
coverage \cite{DetectorsAtlas, DetectorsCMS} and the size of the 
radiative corrections involving the lepton mass 
\cite{Baur98}, which are not considered here anyway. 
A realistic simulation should include the detector 
acceptances and efficiencies. 
We imitate the experimental set up at LHC (Tevatron) 
smearing the lepton and jet energies using values based on the CDF 
specifications \cite{Cuts} 
\be
\frac{\Delta E^e}{E^e}=\frac{10 (20) \%}{\sqrt {E^e}}
+ 0.3 (2) \%, \,\,\, 
\frac{\Delta E^j}{E^j}=\frac{50 (80) \%}{\sqrt {E^j}}
+ 3 (5) \% ,
\label{smearing}
\en 
with $E$ in GeV, and requiring 
that the momenta $p$, pseudorapidities $\eta$ and separation in 
the pseudorapidity - azimuthal angle plane $\Delta R$ satisfy
\bea
& p^e_t=\sqrt{p^{e\,2}_T}>20 \,{\rm GeV}, \,\,\, &
p^j_t=\sqrt{p^{j\,2}_T}>50\, (30) \,{\rm GeV}, \,\,\, 
\nonumber \\
& |\eta ^{e,j}| <2.5, \,\,\, &
\, \, \, \Delta R_{e,j} >0.4, 
\label{cuts}
\ena 
respectively, unless otherwise stated.
In Figure 1 (a) we plot the 
$pp\rightarrow Vj\rightarrow e^-e^+j$ cross section, 
with $V = \gamma , Z$ and the cuts above for LHC, 
as function of 
$M_{e^-e^+}=\sqrt{(p^{e^-}+p^{e^+})^2}$ (upper curves).
The distributions with (solid) and without (dashed) smearing 
are overimposed, no difference being apparent.
In Figure 2 (a) we show the corresponding charge 
asymmetries, ${\rm A}_{\rm FB}$ relative to the initial parton 
and ${\rm A}_{\rm FB}^j$ to the final jet. 
Both give similar results, although the former is adapted to the 
$q\bar q$ collisions and the latter to the
$g\stackrel {(-)}{q}$ ones.
We do not include hadronization neither detector simulation
which, as the smearing, mainly affect the asymmetries, 
in particular
due to the fact that the directions of the jets are related
but not equal to the directions of the parent partons.
In the Figures we also show the 
$pp\rightarrow V\stackrel {(-)}{b}\rightarrow e^-e^+
\stackrel {(-)}{b}$ cross section, 
assuming a $\stackrel {(-)}{b}$-tagging efficiency of 
$50\ \%$ \cite{DetectorsAtlas}, 
and the corresponding asymmetry ${\rm A}_{\rm FB}^b$, 
assuming no charge misassignment 
(thick lines). 
The cross section is a factor 30 smaller in this case, 
but the asymmetry is much larger because only 
$g\stackrel {(-)}{b}$ collisions contribute.
As explained NLO corrections are not included 
but they are eventually needed to describe the data. 
In Figure 1 we also plot the 
top pair background, $p\stackrel {(-)}{p} \rightarrow 
t\stackrel {-}{t} \rightarrow W^+ W^- b \stackrel {-} {b} 
\rightarrow  \nu_e \stackrel {-} {\nu}_e e^+ e^- b 
\stackrel {-} {b} $ \cite{Kleiss}, and consider the case 
of losing one $b$. We assume 
the same $\stackrel {(-)}{b}$-tagging efficiency 
and that the second $b$ jet is missed if 
$p^b_t < 50$ GeV. We also require that the 
total transverse momentum $p_t < 20$ GeV, 
$p = p^{e^-} + p^{e^+} + p^b$.
The resulting distribution is rather flat and 
the smearing makes no difference.
In the $M_{e^-e^+}$ interval between 75 and 105 GeV 
the signal is 200 
times larger, $\sigma ^{Vb} = 1.7$ pb 
whereas $\sigma ^{t\bar t} = 0.008 $ pb. 
This background is further reduced by a factor 1.25 
if the $b$ jet is only missed for $p_t^b < 20\ $GeV. 
So, we neglect it in the following. 
In any case its mixed $e\mu$ decays
can also provide a further handle on $t\bar t$.
In Figures 1 and 2 (b) we plot the same 
cross sections and asymmetries but for Tevatron.
At 2 TeV the $q\bar q$ collisions dominate 
and the asymmetry adapted to these events 
${\rm A}_{\rm FB}$ is much larger. 
The applied smearing and cuts are given in Eqs. 
(\ref{smearing},\ref{cuts}).
In particular, for the $t\bar t$ background 
we mimic the missed  $\stackrel {(-)}{b}$ by demanding
$p^b_t < 30$ GeV and also require
$p_t < 20$ GeV. In such conditions we find that the
$Vb$ signal is 700 times larger in the 
$M_{e^-e^+}$ range between 75 and 105 GeV, 
$\sigma ^{Vb} = 58$ fb 
whereas $\sigma ^{t\bar t} = 0.08$ fb.
Other $W$ pair backgrounds like 
$p\stackrel {(-)}{p} \rightarrow W^+ W^- j, W^+ W^- \stackrel {(-)}{b} $ or 
$W^+ W^- jj , W^+ W^- b \stackrel {-}{b}$ with only one jet 
detected, which can be large a priori, can be further reduced 
requiring small total transverse momentum. 

Near the $Z$ pole, $M_{e^-e^+}\sim M_Z$,  
the asymmetries can be approximated 
by \cite{Rosner89}
\be
{\rm A} =
{\rm b}({\rm a}-\sin ^2 \theta ^{\rm lept}_{\rm eff} (M_Z^2)),
\label{appr.asymm}
\en 
translating then their measurement into a precise 
determination of 
$\sin ^2 \theta ^{\rm lept}_{\rm eff} (M_Z^2)$. 
In the Table we collect the asymmetry estimates and their 
statistical precision, the corresponding b and a 
values in Eq. (\ref{appr.asymm}) and the 
precision reach 
$\delta \sin ^2 \theta _{\rm eff}^{\rm lept}$ of LHC 
and Tevatron for $M_{e^-e^+}$ in the range $[75,105]\, {\rm GeV}$ 
and two sets of cuts. 
The first set has been used 
throughout the paper and is given in Eq. (\ref{cuts}), 
whereas the second one requires a smaller minimum jet transverse 
momentum, $p_t^j > 20\ (10)$ GeV at LHC (Tevatron). 
These less stringent cuts increase the number 
of events, and then improve the statistical precision by 
10 to 50 $\%$ depending on the asymmetry and collider. 
We have not tried to optimize them at this stage, but it 
will have to be done when dealing with real data and 
the experimental inefficiencies are known. 
The cross sections are also gathered in the Table. 
All the estimates  
include the smearing in Eq. (\ref{smearing}). 
The results without smearing are very similar, except 
for the ${\rm A}_{\rm FB}^j$ asymmetry and the second 
set of cuts for which ${\rm A}_{\rm FB}^j$ is 20 $\%$ 
smaller (larger) at LHC (Tevatron). 

We have assumed throughout the paper a $\stackrel {(-)}{b}$-tagging 
efficiency $\epsilon$ of $50\ \%$. This is too 
optimistic, especially because we assume no contamination 
$\omega$, and in particular no charge misidentification. 
The statistical precisions $\delta {\rm A}$ and
$\delta \sin ^2 \theta _{\rm eff}^{\rm lept}$ 
are proportional to $\epsilon ^{-\frac{1}{2}}$, 
and the asymmetries A and coefficients b in Eq. (\ref{appr.asymm}) 
to $1-2\omega$. 
This means in particular that the contamination multiplies 
$\delta \sin ^2 \theta _{\rm eff}^{\rm lept}$ 
by $(1-2\omega)^{-1}$.
Hence, if we only consider semileptonic $b$ decays, 
implying $\epsilon \sim 0.1$ and $\omega \sim 0$, 
$\delta {\rm A}$ and
$\delta \sin ^2 \theta _{\rm eff}^{\rm lept}$ increase 
by a factor $\sim 2$.  
In practice we must try to maximize the quality factor 
${\rm Q} = \epsilon (1-2\omega)^2$ \cite{BaBar}. 
The statistical precisions given in the Table are 
certainly optimistic for systematic errors 
are also sizeable. To approach the quoted precisions 
will be an experimental challenge.

In summary, we have pointed out that the large $Vj$ production 
cross section at hadron colliders and the possibility of 
measuring the lepton asymmetries relative to the final jet 
allow for a precise determination of the effective electroweak 
mixing angle. If there is an efficient $b$-tagging and charge 
identification, these events with a $b$ jet also allow for 
a new determination of ${\rm A}_{\rm FB}^b$. 
The corresponding statistical precisions are collected
in the Table. As in Drell-Yan production \cite{Rosner96}, 
this process is also sensitive to new physics for 
large $M_{e^-e^+}$, especially to new gauge bosons.  

\vskip.2cm
\noindent {\bf Acknowledgments}\\
We thank J.A. Aguilar Saavedra, A. Bueno, R. Pittau 
and J. Santiago for useful comments.
This work was supported in part by MCYT under 
contract FPA2000-1558, Junta de Andaluc{\'\i}a group FQM 101 
and the European Community's Human Potential Programme under 
contract HPRN-CT-2000-00149 Physics at Colliders.

\newpage
~
\begin{figure}[ht]
\epsfig{file=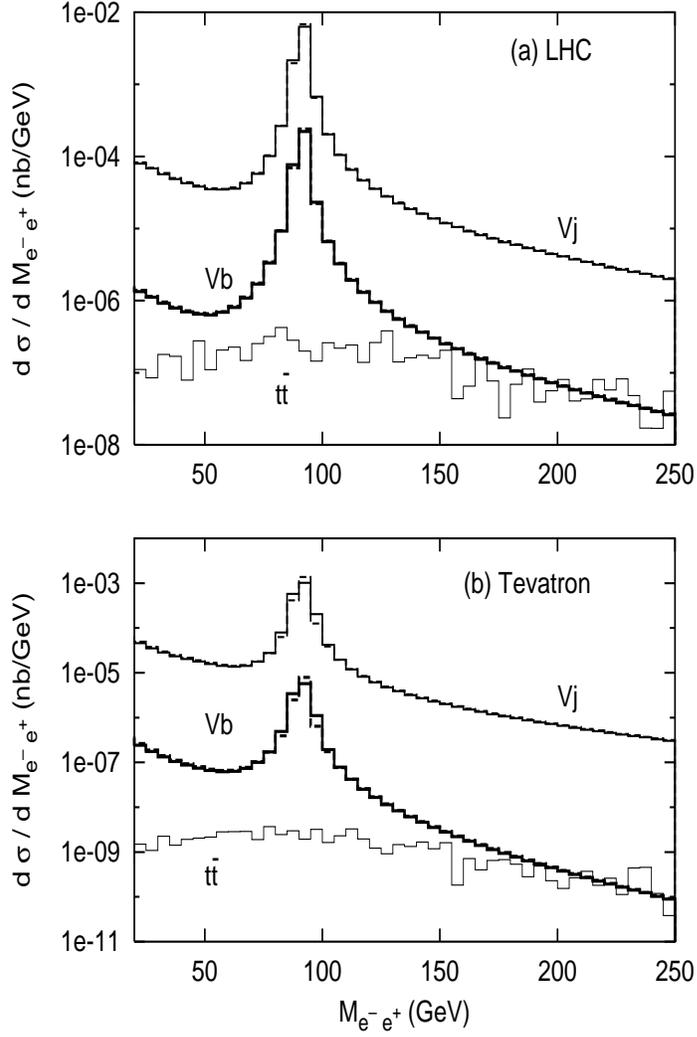,width=14cm,height=16cm,angle=-90}
\caption{\label{crosssection}
Leading order $e^-e^+j$ ($Vj$) and 
$e^-e^+\stackrel{(-)}{b}$ ($Vb$ and $t\bar t$) 
cross sections  
as function of $M_{e^-e^+}$ 
for the processes, cuts 
and efficiencies discussed in the text 
at LHC (a) and Tevatron (b).}
\end{figure}

\newpage
~

\begin{figure}[ht]
\epsfig{file=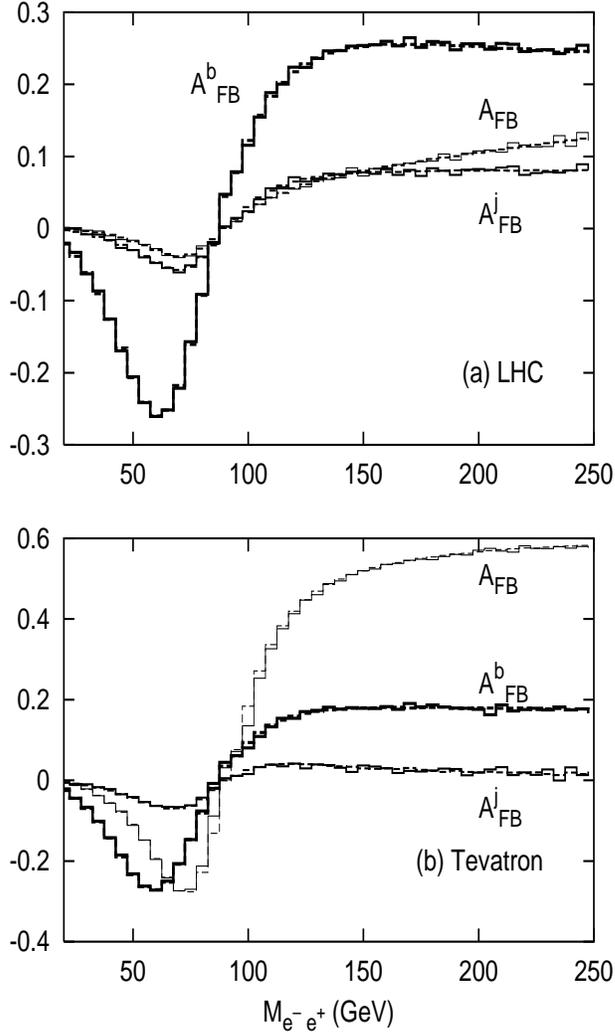,width=14cm,height=16cm,angle=-90}
\caption{\label{transverse}
Forward-backward electron asymmetries 
defined in the text as function of $M_{e^-e^+}$ 
for $e^-e^+j$ (${\rm A}_{\rm FB}$ and 
${\rm A}_{\rm FB}^j$) and $e^-e^+\stackrel{(-)}{b}$ 
(${\rm A}_{\rm FB}^b$) events in Figure 1 
at LHC (a) and Tevatron (b). The $t\bar t$ background 
is not included.}
\end{figure}

\newpage
~ 

\begin{table*}[ht]
\begin{center}
\begin{tabular}{|c|c|c||c|c|c|c|c|}
\hline
& $\sigma$ (pb)& & A & $\delta {\rm A}$ & b & a  &
$\delta \sin^2\theta_{\rm eff}^{\rm lept} $ \\
\hline
\hline
LHC & $\sigma^{Vj} = 49$  & $A_{FB}$  & $8.708\times 10^{-3}$  &
$4.5\times 10^{-4}$  
&$0.346$ 
&$0.2491$ 
&$1.3\times 10^{-3}$ \\
$p_t^j > 50$~GeV & & $A_{FB}^j$&  $1.170\times 10^{-2}$  &$4.5\times 10^{-4}$ 
& $0.467$ 
& $0.2490$ 
& $9.7\times 10^{-4}$\\
& $\sigma^{Vb} = 1.7$ & $A_{FB}^b$  &$7.136\times 10^{-2}$   &
$2.4\times 10^{-3}$ 
& $2.723$
&$0.2502$ 
&$8.9\times 10^{-4}$  \\
\hline

& $\sigma^{Vj} = 167$ & $A_{FB}$  &$8.207\times 10^{-3}$ & 
$2.4\times 10^{-4}$
& $0.357$ 
&$0.2469$ 
& $6.9\times 10^{-4}$ \\
$p_t^j > 20$~GeV & & $A_{FB}^j$&   $8.077\times 10^{-3}$  & 
$2.4\times 10^{-4}$ & 
$0.289$ & 
$0.2519$ & 
$8.5\times 10^{-4}$\\
&$\sigma^{Vb} = 5.9$ & $A_{FB}^b$  & $5.667\times 10^{-2}$ & 
$1.3\times 10^{-3}$& 
$2.187$ & 
$0.2499$ & 
$6.0\times 10^{-4}$\\
\hline
Tevatron& $\sigma^{Vj} = 9.7$ & $A_{FB}$ & $5.944\times 10^{-2}$ &
$3.2\times 10^{-3}$ &
$2.658$ &
$0.2463$ &
$1.2\times 10^{-3}$ \\
$p_t^j > 30$~GeV & &  $A_{FB}^j$ & $8.306\times 10^{-3}$  &
$3.2\times 10^{-3}$ & 
$0.386$& 
$0.2455$ & 
$8.3\times 10^{-3}$ \\
&$\sigma^{Vb} = 0.06$ &  $A_{FB}^b$  &$5.373\times 10^{-2}$ &
$4.2\times 10^{-2}$ &
$2.206$ & 
$0.2483$ &
$1.9\times 10^{-2}$\\
\hline
& $\sigma^{Vj} = 39$ & ${\rm A}_{\rm FB}$ & $6.722\times 10^{-2}$ &
$1.6\times 10^{-3}$ &
$3.005$ &
$0.2463$ &
$5.3\times 10^{-4}$ \\
$p_t^j > 10$~GeV &  & ${\rm A}_{\rm FB}^j$ & $6.374\times 10^{-3}$  &
$1.6\times 10^{-3}$ & 
$0.357$& 
$0.2418$ & 
$4.5\times 10^{-3}$ \\
& $\sigma^{Vb} = 0.21$ & ${\rm A}_{\rm FB}^b$ &$4.709\times 10^{-2}$ &
$2.2\times 10^{-2}$ &
$1.924$ & 
$0.2484$ & 
$1.1\times 10^{-2}$\\
\hline
\end{tabular}
\end{center}
\caption{ 
Estimates for the $e^-e^+j$ and $e^-e^+\stackrel {(-)}{b}$ 
cross sections and asymmetries defined in the text
with $M_{e^-e^+}$ in the range $[75,105]\, {\rm GeV}$. The 
statistical precisions are also given.
The integrated luminosity as well as 
the smearing, cuts and tagging efficiency can be found
in the text.}
\label{table1}
\end{table*}     

\vfill

\end{document}